\documentclass[fleqn,twoside]{article}
\usepackage{espcrc2}

\readRCS
$Id: espcrc2.tex,v 1.2 2004/02/24 11:22:11 spepping Exp $
\usepackage{epsfig}

\newcommand{\AmS}{{\protect\the\textfont2
  A\kern-.1667em\lower.5ex\hbox{M}\kern-.125emS}}

\def\bom#1{{\mbox{\boldmath $#1$}}}
\def\la{\langle}
\def\ra{\rangle}
\def\ket#1{|{#1}\ra}

\def\cm{{\cal M}}
\def\sp{{\bom {Sp}}}
\def\nn{\nonumber}
\def\ep{\epsilon}

\def\as{\alpha_{\mathrm{S}}}
\def\gs{g_{\mathrm{S}}}

\def\Eq#1{Eq.~({\ref{#1}})}

\def\beq{\begin{equation}}
\def\eeq{\end{equation}}
\def\bea{\begin{eqnarray}}
\def\eea{\end{eqnarray}}

\title{\vspace{-2cm}
\hfill \parbox{3cm}{\normalsize \tt IFIC/04-32 \\ RBRC-422 \\ BNL-NT-04/22} \vspace{1cm} \\
Collinear splitting, parton evolution and the strange-quark asymmetry of the
nucleon in NNLO QCD\thanks{Talk presented by G. Rodrigo.}}

\author{Germ\'an Rodrigo\address{Instituto de F\'{\i}sica Corpuscular, 
        Apartado de Correos 22085, E-46071 Valencia, Spain}\thanks{E-mail: 
        {\tt german.rodrigo@ific.uv.es}.
        Supported by Generalitat Valenciana under grant 
        GRUPOS03/013, and MCyT under grant FPA-2001-3031.},
        Stefano Catani\address{INFN, Sezione di Firenze and
        Dipartimento di Fisica, Universit\`a di Firenze, \\
        I-50019 Sesto Fiorentino, Florence, Italy}\thanks{
        E-mail: {\tt stefano.catani@fi.infn.it}},
        Daniel de Florian\address{Departamento de F\'\i sica, FCEYN, 
        Universidad de Buenos Aires,
        Argentina}\thanks{E-mail: {\tt deflo@df.uba.ar}. Supported by 
        Conicet, Fundaci\'on Antorchas, UBACyT and ANPCyT.},
        and
        Werner Vogelsang\address{Physics Department and RIKEN-BNL 
        Research Center, Brookhaven National Laboratory, \\ 
        Upton, NY 11973, U.S.A.}\thanks{E-mail: 
        {\tt wvogelsang@bnl.gov}. Supported by 
        the U.S.\ Department of Energy (contract number DE-AC02-98CH10886).}}

\begin{document}

\begin{abstract}
We consider the collinear limit of QCD amplitudes at one-loop 
order, and their factorization properties directly in colour space. 
These results apply to the multiple collinear limit of an arbitrary 
number of QCD partons, and are a basic ingredient in many higher-order 
computations. In particular, we discuss the triple collinear limit 
and its relation to flavour asymmetries in the QCD evolution of parton 
densities at three loops. As a phenomenological consequence of 
this new effect, and of the fact that the nucleon has non-vanishing 
quark valence densities, we study the perturbative generation of a 
strange--antistrange asymmetry $s(x)-\bar{s}(x)$ in the nucleon's sea. 
\vspace*{-4.5mm}
\end{abstract}

\maketitle

\section{INTRODUCTION}
\vspace*{-1mm}
The high precision of experiments at past, present and future particle 
colliders (LEP, HERA, Tevatron, LHC, $e^+e^-$ linear colliders) demands
a corresponding precision in theoretical predictions.
As for perturbative QCD predictions, this means calculations beyond
the next-to-leading order (NLO) in the strong coupling $\as$.
Recent years have witnessed much progress in this field.
In particular, a great deal of work has been devoted to study the
properties of QCD scattering amplitudes in the infrared (soft and collinear)
region~\cite{Berends:1988zn}--\cite{Badger:2004uk}.

The understanding of the infrared singular behaviour of 
QCD amplitudes is a prerequisite for the evaluation of infrared-finite 
cross sections (and, more generally, infrared- and collinear-safe QCD 
observables) at higher orders in perturbation theory.
Moreover,
the information on the infrared properties of the amplitudes can be 
exploited to compute large (logarithmically enhanced) perturbative terms
and to resum them to all perturbative orders~\cite{resex}.
Those investigations are also valuable for improving the 
physics content of Monte Carlo event generators
(see e.g. Ref.~\cite{mcws}). In addition, these studies
prove to be useful even beyond the strict QCD context,
and can provide hints on the structure of highly symmetric gauge 
theories at infinite orders in the perturbative expansion
(e.g. N=4 super-Yang-Mills, see Ref.~\cite{Anastasiou:2003kj}).

Another important application is the calculation of the Altarelli--Parisi 
(AP) kernels, that control the scale evolution of parton densities and 
fragmentation functions. The calculation of the 
next-to-next-to-leading order (NNLO)
kernels has been completed very recently~\cite{Moch:2004pa}. 
Collinear factorization at the amplitude level (see Sect.~\ref{sec:collinear})
can be used~\cite{Kosower:2003np} 
as an alternative and independent method 
to perform that calculation.
To this purpose, the one-loop triple collinear splitting~\cite{Catani:2003vu}
(see Sect.~\ref{sec:triple})
is one of the necessary ingredients. Two other ingredients are
the tree-level quadruple collinear splitting~\cite{DelDuca:1999ha} 
and the two-loop double collinear splitting~\cite{Badger:2004uk}.

Besides increasing the quantitative precision of the theoretical calculations,
the evaluation of higher-order contributions can reveal qualitatively new
quantum effects.
An interesting example is the 
perturbative generation of charge asymmetries in the nucleon's 
sea~\cite{Catani:2004nc} (see Sect.~\ref{sec:asym}),
which arises from the NNLO evolution 
of parton densities. 

\vspace*{-3mm}
\section{COLLINEAR FACTORIZATION IN COLOUR SPACE}
\label{sec:collinear}
\vspace*{-1mm}
We consider a generic scattering process involving
final-state QCD partons (massless quarks and gluons)
of flavour $a_1,a_2,\dots$ and momenta $p_1, p_2, \dots$,
which is described by the matrix element 
$\cm_{a_1,a_2,\dots}(p_1,p_2,\dots)$; the external 
legs are on shell $(p_i^2=0)$ and have physical spin polarizations.
Up to one-loop order, one can write 
\vspace{-1mm}
\beq
\label{loopex}
\cm = \left( \gs \right)^q \; \left[ \; \cm^{(0)}
+ \frac{\as}{2\pi} \;\cm^{(1)}
+ {\cal O}(\as^2) \right]~,
\eeq
where the overall power $q$ is integer. The one-loop amplitude $\cm^{(1)}$ 
contains ultraviolet and infrared singularities that
are regularized by using dimensional regularization 
in $d=4- 2\ep$ space-time dimensions
($\mu$ is the dimensional-regularization scale).

The multiple collinear limit is approached when the momenta 
$p_1, \dots, p_m$ of $m$ partons become parallel.
In this limit all the particle subenergies $s_{ij}=(p_i+p_j)^2$, with
$i,j=1,\dots,m$, are of the same order and vanish simultaneously, 
and the matrix element $\cm(p_1,\dots,p_m,p_{m+1},\dots)$
becomes singular. At the tree level, the dominant singular behaviour is
$\cm^{(0)}(p_1,\dots,p_m,p_{m+1},\dots)\sim (1/{\sqrt s})^{m-1}$,
where $s$ generically denotes a two-particle subenergy $s_{ij}$, 
or a three-particle subenergy $s_{ijk}$, and so forth. At one-loop order, 
this singular behaviour is simply modified by scaling violation,
$\cm^{(1)}(p_1,\dots,p_m,p_{m+1},\dots)\sim (1/{\sqrt s})^{m-1} (s/\mu^2)^{-\ep}$.
The dominant singular behaviour is captured by universal (process-independent) 
factorization formulae, that are usually presented upon decomposition in colour 
subamplitudes~\cite{Bern:zx}--\cite{Kosower:2002su}.
Collinear factorization is nonetheless valid directly in colour 
space~\cite{Catani:2003vu}.

\begin{figure}[tb]
\begin{center}
\vspace{-.3cm}
\epsfig{figure=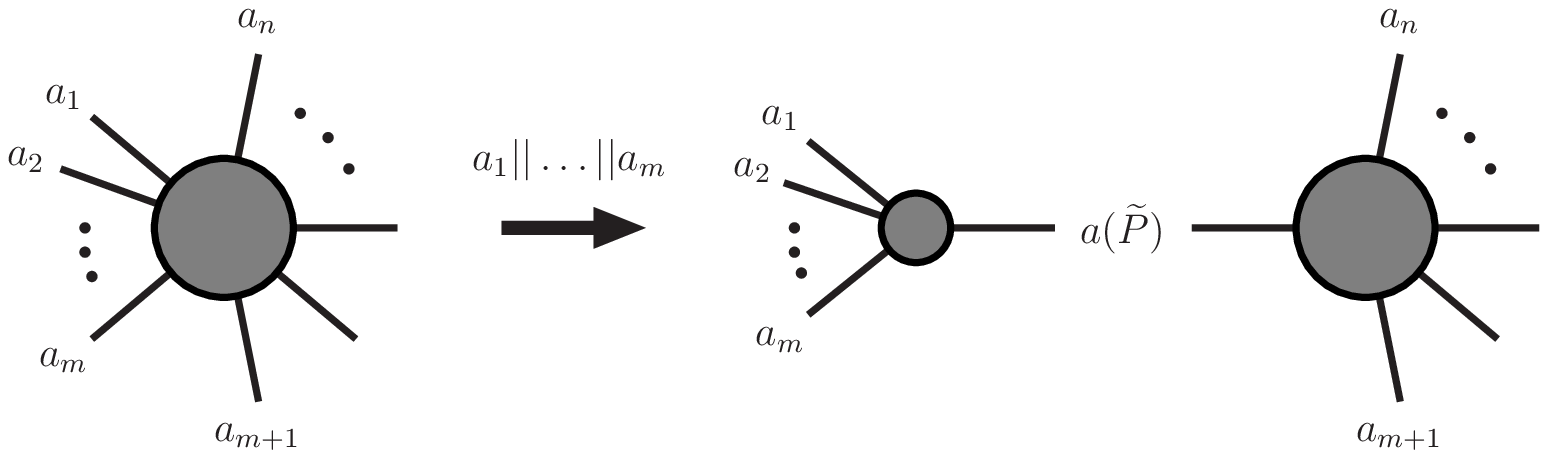,width=0.4\textwidth}
\end{center}
\caption{Factorization of tree-level amplitudes in the multiple collinear
limit. \label{factoriza_lo} \vspace{-.4cm}}
\end{figure}
\begin{figure}[tb]
\begin{center}
\epsfig{figure=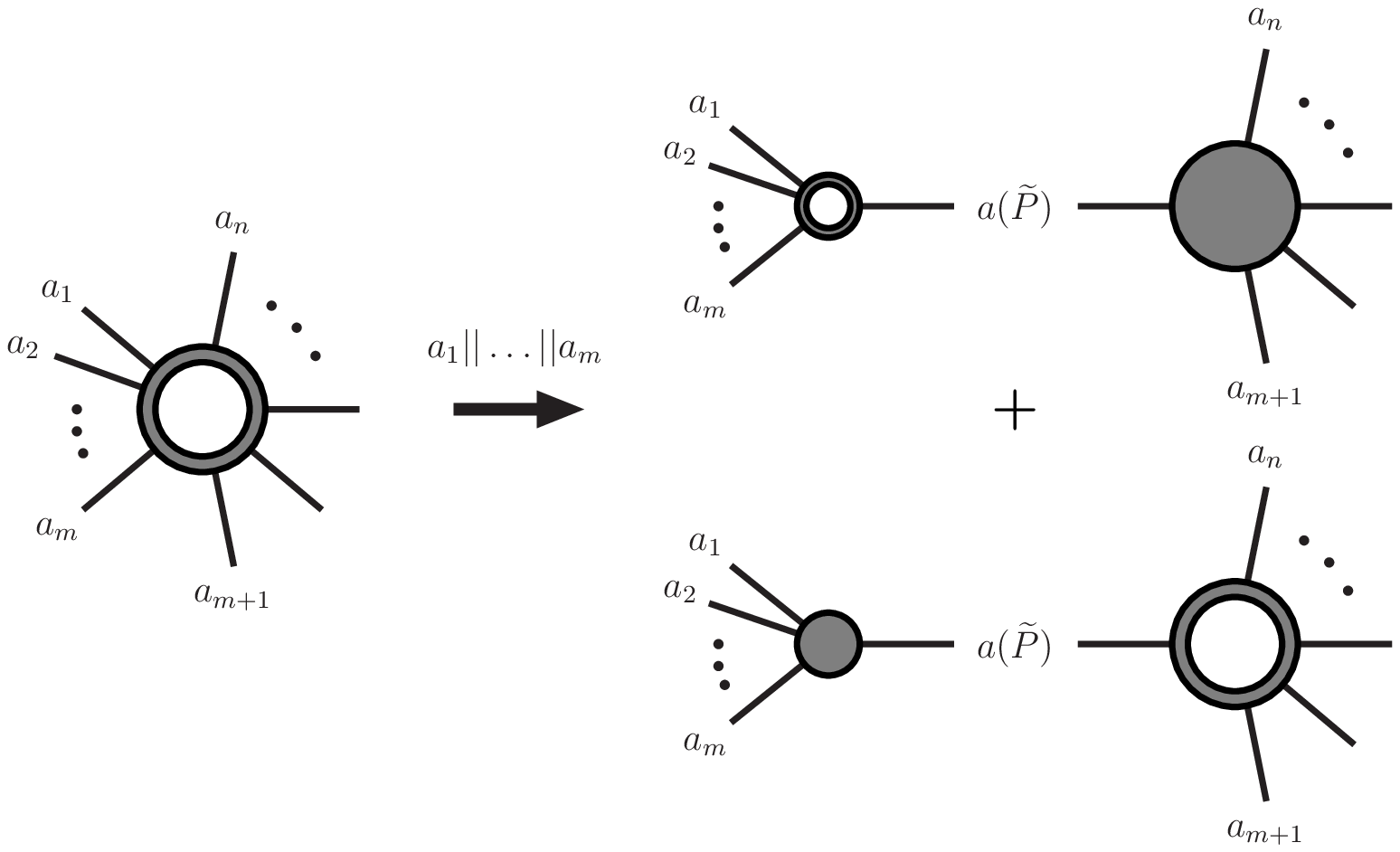,width=0.4\textwidth}
\end{center}
\caption{Factorization of one-loop amplitudes in the multiple collinear 
limit. \label{factoriza_nlo} \vspace{-.4cm}}
\end{figure}

The colour-space factorization formulae for the multiple
collinear limit of the tree-level and one-loop amplitudes $\cm^{(0)}$ 
and $\cm^{(1)}$ are:
\vspace*{-1.5mm}
\bea
\label{facttree}
\!\!\!\!\!\!\!\!\!\!\!\!
&& \ket{\cm_{a_1,\dots,a_m,a_{m+1},\dots}^{(0)}(p_1,\dots,p_m,p_{m+1},\dots)}
\simeq \\ 
\!\!\!\!\!\!\!\!\!\!\!\!
&& \sp_{a_1 \dots a_m}^{(0)}(p_1,\dots,p_m) 
\; \ket{\cm_{a,a_{m+1},\dots}^{(0)}({\widetilde P},p_{m+1},\dots)}~, \nn
\eea
\bea
\label{factone}
\!\!\!\!\!\!\!\!\!\!\!\!
&& \ket{\cm_{a_1,\dots,a_m,a_{m+1},\dots}^{(1)}(p_1,\dots,p_m,p_{m+1}\dots)}
\simeq \\
\!\!\!\!\!\!\!\!\!\!\!\!
&& \sp_{a_1 \dots a_m}^{(1)}(p_1,\dots,p_m) 
\; \ket{\cm_{a,a_{m+1}\dots}^{(0)}({\widetilde P},p_{m+1},\dots)} 
+ \nn \\
\!\!\!\!\!\!\!\!\!\!\!\!
&& \sp_{a_1 \dots a_m}^{(0)}(p_1,\dots,p_m) 
\; \ket{\cm_{a,a_{m+1},\dots}^{(1)}({\widetilde P},p_{m+1},\dots)}~. \nn 
\eea
These factorization formulae are valid in any number $d=4-2\ep$ 
of space-time dimensions.
The only approximation involved on the right-hand sides amounts to 
neglecting terms that are less singular in the multiple collinear limit.
A graphical representation of the factorization formulae is shown 
in Figs.~\ref{factoriza_lo} and~\ref{factoriza_nlo}.
Equations (\ref{facttree}) and (\ref{factone}) relate the original
matrix element (on the left-hand side) with $m+k$ partons 
(where $k$ is arbitrary) to a matrix element (on the right-hand side) 
with $1+k$ partons. The latter is obtained from the former
by replacing the $m$ collinear partons with a single parent parton, 
whose momentum ${\widetilde P}$ (${\widetilde P}^2=0$) defines 
the collinear direction 
and whose flavour $a$ is determined 
by flavour conservation in the splitting process $a \to a_1+ \dots+ a_m$.
The derivation of \Eq{facttree} in colour space is quite 
straightforward~\cite{Catani:1998nv}. Its one-loop 
extension, \Eq{factone}, 
follows,
in particular, 
from colour coherence of QCD radiation~\cite{inprep}. 

The process dependence of the factorization formulae is entirely embodied
in the matrix elements. The tree-level and one-loop factors 
$\sp^{(0)}_{a_1 \cdots a_m}$ and $\sp^{(1)}_{a_1 \cdots a_m}$, which
encode the singular behaviour in the multiple collinear limit,
are universal (process-independent). 
They depend on the momenta and quantum numbers (flavour, spin, colour) 
of the $m$ partons that arise from the collinear splitting. 
The {\em splitting matrix} $\sp_{a_1 \cdots a_m}$ is a matrix in 
colour+spin space, acting onto the colour and spin indices of the 
$m$ collinear partons on the left and onto the colour and spin indices 
of the parent parton on the right.

The square of the splitting matrix $\sp_{a_1 \cdots a_m}$, summed over 
final-state colours and spins and averaged over colours and spins of 
the parent parton, defines the $m$-parton splitting function 
$\langle \hat{P}_{a_1 \cdots a_m} \rangle$, 
which is a generalization of the customary (i.e. with $m=2$)
AP
splitting function~\cite{Altarelli:1977zs}. 
The normalization of the tree-level $\la \hat{P}^{(0)}_{a_1 \cdots a_m} \ra$ 
and one-loop $\la \hat{P}^{(1)}_{a_1 \cdots a_m} \ra$ splitting functions 
is fixed by 
\bea
\label{p0def}
\!\!\!\!\!\!\!\!\!\!\!\!
&&\langle \hat{P}_{a_1 \cdots a_m}^{(0)} \rangle = 
\left( \frac{s_{1 \dots m}}{2 \;\mu^{2\ep}} \right)^{m-1} \;
{\overline {| \sp_{a_1 \dots a_m}^{(0)} |^2}}~, \\
\!\!\!\!\!\!\!\!\!\!\!\!
&& \langle \hat{P}_{a_1 \cdots a_m}^{(1)} \rangle = \nn \\ 
\!\!\!\!\!\!\!\!\!\!\!\! && \;\;\;\;
\left( \frac{s_{1 \dots m}}{2 \;\mu^{2\ep}} \right)^{m-1} 
{\overline {\left[(\sp^{(0)}_{a_1 \cdots a_m})^\dagger \, \sp^{(1)}_{a_1 \cdots a_m} 
+ \mathrm{h.c.} \right]}}~. \nn
\eea

The one-loop amplitude $\cm^{(1)}$ and, hence, 
$\sp^{(1)}$
have ultraviolet and infrared divergences that show up as $\ep$-poles in dimensional 
regularization. The one-loop splitting matrix can be decomposed as
\beq
\label{sp1df}
\sp^{(1)}_{a_1 \cdots a_m} = \sp^{(1) \,{\rm div.}}_{a_1 \cdots a_m} + 
\sp^{(1) \,{\rm fin.}}_{a_1 \cdots a_m}~,
\eeq
where $\sp^{(1) \,{\rm div.}}_{a_1 \cdots a_m}$ 
contains all the $\ep$-poles
and $\sp^{(1) \,{\rm fin.}}_{a_1 \cdots a_m}$ is finite when $\ep \to 0$.
In Ref.~\cite{Catani:2003vu} we have presented the explicit expression
of $\sp^{(1) \,{\rm div.}}_{a_1 \cdots a_m}$
for an arbitrary number $m$ of final-state collinear partons. 


\vspace*{-4mm}
\section{ONE-LOOP TRIPLE COLLINEAR SPLITTING}
\label{sec:triple}
\begin{figure}
\vspace{-2.2cm}
\centerline{
   \epsfig{figure=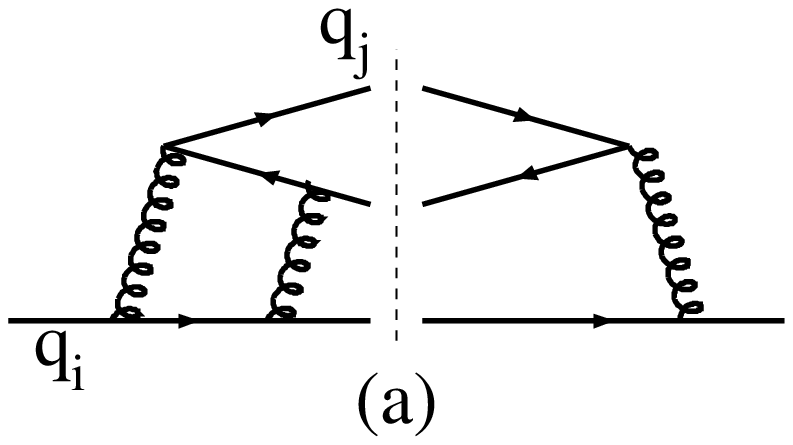,width=0.4\textwidth,clip=}
   \hfill \hspace{-3cm}
   \epsfig{figure=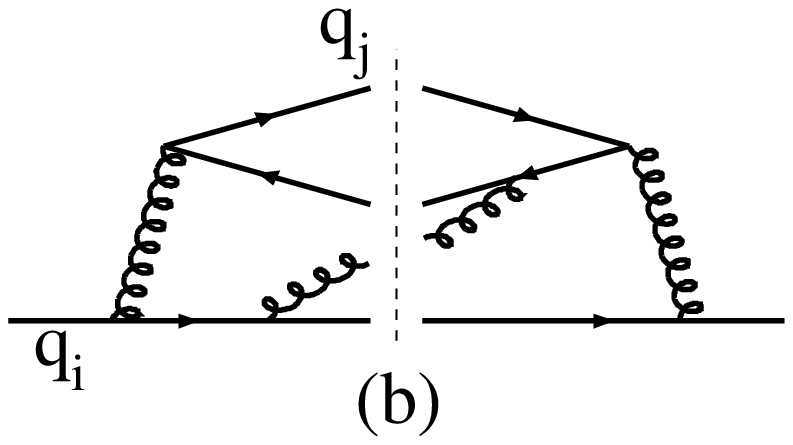,width=0.4\textwidth,clip=} }\vspace{-6.2cm}
\caption{\label{diagrams} Example of (a) virtual and (b)
real contributions to $P_{ns}^{(2) S}$.\vspace{-.4cm}}
\end{figure}

\vspace*{-1mm}
The one-loop splitting 
amplitudes for the double collinear limit $a \to a_1+a_2$
are known \cite{Bern:zx,Bern:1998sc,Kosower:1999xi}.
As a first step beyond the double collinear limit, we have 
considered~\cite{Catani:2003vu} the triple collinear splitting 
process $q \to q+\bar{q}^{\prime}+q^{\prime}$, where $q$ and $q^{\prime}$ 
denote quarks of different flavours. To evaluate the one-loop 
splitting matrix we have used a process-independent 
method~\cite{Catani:1998nv,Kosower:1999xi,inprep}.
Considering physical spin polarizations,
the splitting matrix is calculated from the sole 
Feynman diagrams where the parent parton emits and 
absorbs collinear radiation.
In the case $q \to q+\bar{q}^{\prime}+q^{\prime}$, one has to consider, for
instance, the one-loop  
diagram depicted in Fig.~\ref{diagrams}(a). 

Our computation of $\sp^{(1)}$ requires the evaluation of a set of basic one-loop
(scalar and tensor) integrals. Besides the customary one-loop 
integrals, new integrals with additional propagators of the 
type $1/(n\cdot q)$
($q$ is the loop momentum and $n$ is an auxiliary 
light-like vector),
which come from the physical polarizations of 
the virtual gluons,  have to be calculated.
Some of these integrals, which resemble those encountered in axial-gauge
calculations, were evaluated 
in Ref.~\cite{Kosower:1999xi}
in the context of 
the
calculation of the one-loop double collinear 
splitting $a \to a_1+a_2$. More complicated 
integrals (higher-point functions) of this type are involved in 
triple collinear splitting processes. We have computed 
(to high orders in the $\ep$ expansion) all the basic one-loop 
integrals that appear in any triple collinear 
splitting. These results can be applied 
to evaluate the one-loop splitting matrix of 
any splitting process $a\to a_1+a_2+a_3$~\cite{inprep}.

The explicit expressions up to ${\cal O}(\ep^0)$ of the splitting matrix 
$\sp_{q \bar{q}^\prime q^\prime}$ and of the corresponding splitting 
function 
$\la \hat{P}^{(1)}_{q \bar{q}^\prime q^\prime} \ra$ 
(see Fig.~\ref{diagrams}(a))
are presented in
Ref.~\cite{Catani:2003vu}. 
It is important to observe that 
$\la \hat{P}^{(1)}_{q \bar{q}^\prime q^\prime} \ra$ has a contribution
(which is proportional to the color factor $d^{abc} d_{abc}$)
that changes sign by exchanging the momenta of the evolved quark and antiquark 
$q^\prime$ and $\bar{q}^\prime$. 
This charge asymmetry is a new quantum effect
produced by the exchange of three gluons in the $t$-channel.
When the charge asymmetry of 
$\la \hat{P}^{(1)}_{q \bar{q}^\prime q^\prime} \ra$ is combined with the
corresponding tree-level contribution (see Fig.~\ref{diagrams}(b)), 
it leads to a non-vanishing value of the NNLO 
AP
kernel 
$P_{qq^\prime} - P_{q\bar{q}^\prime}$. The main physical consequence of this
effect is discussed in Sect.~\ref{sec:asym}.

\vspace*{-3mm}
\section{STRANGE-QUARK ASYMMETRY IN THE NUCLEON}
\label{sec:asym}
\vspace*{-1mm}
Strange quarks and antiquarks play a fundamental
role in the structure of the nucleon~\cite{ellis}.
Among the various strangeness--related properties of the
nucleon, the strange ``asymmetry'', $s(x)-\bar{s}(x)$,
in the densities of strange quarks and antiquarks,
$x$ being the light-cone momentum fraction
they carry, is of particular interest. Since the nucleon
does not carry any strangeness quantum number, the integral
of the asymmetry over all values of $x$ has to vanish:
\vspace*{-1mm}
\begin{equation}
\la s-\bar{s}\ra \equiv \int_0^1 dx \, \left[ s(x)- \bar{s}(x)\right]=0 \; .
\end{equation}
However, there is no symmetry that would prevent the
$x$ dependences of the functions $s(x)$ and $\bar{s}(x)$
from being different. Therefore one can expect
$s(x)\neq  \bar{s}(x)$, in general.

Strange--antistrange asymmetries have been extensively 
discussed in the literature. 
Various non-perturbative
models of the nucleon 
structure~\cite{Cao:1999da} predict a fairly small value of the 
second moment of the strange--antistrange 
distribution,  $|\la x(s-\bar{s})\ra|\sim 10^{-4}$.
A global analysis of unpolarized parton distributions~\cite{Barone:1999yv} 
reported improvements in the data fit 
if the asymmetry $s(x)-\bar{s}(x)$ is positive at high $x$.
However, a recent update~\cite{Portheault:2004xy} of this 
analysis reduces the asymmetry significantly.
The most recent global QCD fit~\cite{Olness:2003wz}
finds a large uncertainty for that asymmetry and quotes
a range $-0.001 < \la x(s-\bar{s})\ra < 0.004$.
The strange asymmetry in the nucleon has become particularly 
relevant in view of the ``anomaly'' seen by the
NuTeV collaboration in their measurement of the Weinberg 
angle~\cite{Zeller:2001hh}. The anomaly could be  
partly explained~\cite{Davidson:2001ji,Kretzer:2003wy}
by a positive value of the second moment $\la x(s-\bar{s})\ra$.

The discussion reported so far regards strange--antistrange
asymmetries that are generated by non-perturbative mechanisms.
Then, because of the customary scaling violation, the asymmetry 
becomes dependent on the hard-scattering scale $Q$ at which
the nucleon is probed.

Perturbative QCD alone, however, definitely predicts a 
non-vanishing and $Q$-dependent value of the strange--antistrange 
asymmetry~\cite{Catani:2004nc}. The effect arises because at NNLO 
in perturbation theory the probability of the inclusive collinear 
splitting (evolution) $q\to q'$ 
becomes different from that of $q\to \bar{q}'$, and because
the nucleon has $u$ and $d$ valence densities. 

Owing to charge conjugation invariance and flavour symmetry of QCD,
the AP kernels that control the parton
evolution of quark--antiquark asymmetries
can be written as
(see e.g. Ref.~\cite{fp})
\vspace*{-5mm}
\bea
P_{q_i q_j}-P_{q_i \bar{q}_j} &=& P_{\bar{q}_i \bar{q}_j} - P_{\bar{q}_i q_j}
\nn \\
&=& \delta_{ij} \;P^{(-)} + ( P_{qq^\prime}-P_{q\bar{q}^\prime} ) \;.
\eea
The AP kernels $P_{qq^\prime}$ and $P_{q\bar{q}^\prime}$ describe
splittings in which the flavor of the quark changes, and
$P_{qq^\prime} \neq P_{q\bar{q}^\prime}$ starting from NNLO \cite{fp,Catani:1994sq}.
In particular, as discussed at the end of Sect.~\ref{sec:triple},
$P_{qq^\prime} - P_{q\bar{q}^\prime}
=(\as/(4\pi))^3 P_{ns}^{(2) S}/N_f$, because of the charge asymmetry produced
by quantum effects at order $\as^3$. 
The explicit expression of 
$P_{ns}^{(2) S}$ is now available thanks to the recent computation 
by Moch, Vermaseren and Vogt~\cite{Moch:2004pa}.

The solution of the AP
equations for the evolution (between the scales $Q_0$ and $Q$)
of the $N$-moment, $(s - {\bar s})_N=\la x^{N-1}(s - {\bar s}) \ra$, 
of the strange-quark asymmetry
reads~\cite{Catani:2004nc}
\vspace*{-1mm}
\bea
\label{sasym}
&& \!\!\!\!\!\!\!\!\!\!\!\!
\left( s - {\bar s} \right)_N(Q^2)=
U_N(Q,Q_0) \; \big[ \; \left( s - {\bar s} \right)_N(Q_0^2) \nn \\ &&
+ \,\delta P_N^{(2)} \left(\as^2(Q) - \as^2(Q_0) \right)
 q^{(V)}_N(Q_0^2) \;\big]~,
\eea
where $q^{(V)} \equiv \sum_{i=1}^{N_f} ( q_i - {\bar q}_i)$ is 
the valence density of the nucleon, 
$U_N$ is the evolution operator controlled by 
$P^{(-)}$,
and $\delta P_N^{(2)}$ is proportional to the NNLO kernel 
$P_{ns}^{(2) S}$~\cite{Catani:2004nc}.
At LO and NLO, $\delta P_N^{(2)}=0$,
and thus any asymmetry
can only be produced by a corresponding asymmetry at the scale 
$Q_0$. Starting from NNLO, the degeneracy of $P_{qq^\prime}$
and $P_{q\bar{q}^\prime}$ is removed, and perturbative QCD necessarily 
predicts a non-vanishing $s - {\bar s}$ asymmetry 
driven by the valence density. 

Predictions for $(s - {\bar s})(x,Q^2)$ based on 
\Eq{sasym} have been presented in Ref.~\cite{Catani:2004nc}.
The asymmetry, $s - {\bar s}$, is set to zero at a given low scale $Q_0$,
and then evolved upwards.
The generated asymmetry is fairly sizable and turns out to
be positive at small $x$ and negative at large $x$. 
Using $Q_0 \sim 0.5$~GeV (as in the `radiative' parton model analysis
of Ref.~\cite{Gluck:1998xa}),
a negative second moment is found:
\vspace*{-1mm}
\begin{equation} \label{secev}
\la x(s-\bar{s})\ra \approx -5\times 10^{-4} \;\;\;\;\;\;
(Q^2=20\,\mathrm{GeV}^2)~.
\end{equation}
\vspace*{-1mm}
The analysis has also be extended~\cite{Catani:2004nc} to predict
the asymmetries of heavy flavours $c$ and $b$. 

{\it Acknowledgments:} G.R. thanks S.~Moch and J.~Bl\"umlein for 
their kind invitation to present these results at 
Zinnowitz, and for the pleasant organization of the workshop.


\vspace*{-2mm}
\begin{thebibliography}{9}
\vspace*{-2mm}

\bibitem{Berends:1988zn}
F.~A.~Berends and W.~T.~Giele,
Nucl.\ Phys.\ B {\bf 313} (1989) 595.

\bibitem{Catani:1998bh}
S.~Catani,
Phys.\ Lett.\ B {\bf 427} (1998) 161;
S.~Catani and M.~Grazzini,
Nucl.\ Phys.\ B {\bf 591} (2000) 435.

\bibitem{Campbell:1997hg}
J.~M.~Campbell and E.~W.~Glover,
Nucl.\ Phys.\ B {\bf 527} (1998) 264.

\bibitem{Catani:1998nv}
S.~Catani and M.~Grazzini,
Phys.\ Lett.\ B {\bf 446} (1999) 143,
Nucl.\ Phys.\ B {\bf 570} (2000) 287.

\bibitem{Bern:zx}
Z.~Bern, L.~J.~Dixon, D.~C.~Dunbar and D.~A.~Kosower,
Nucl.\ Phys.\ B {\bf 425} (1994) 217.

\bibitem{Bern:1998sc}
Z.~Bern, V.~Del Duca and C.~R.~Schmidt,
Phys.\ Lett.\ B {\bf 445} (1998) 168;
Z.~Bern, V.~Del Duca, W.~B.~Kilgore and C.~R.~Schmidt,
Phys.\ Rev.\ D {\bf 60} (1999) 116001.

\bibitem{Kosower:1999xi}
D.~A.~Kosower,
Nucl.\ Phys.\ B {\bf 552} (1999) 319;
D.~A.~Kosower and P.~Uwer,
Nucl.\ Phys.\ B {\bf 563} (1999) 477.

\bibitem{DelDuca:1999ha}
V.~Del Duca, A.~Frizzo and F.~Maltoni,
Nucl.\ Phys.\ B {\bf 568} (2000) 211.

\bibitem{Kosower:2002su}
D.~A.~Kosower,
Phys.\ Rev.\ D {\bf 67} (2003) 116003,
Phys.\ Rev.\ Lett.\  {\bf 91} (2003) 061602,
hep-ph/0311272.

\bibitem{Sterman:2002qn}
G.~Sterman and M.~E.~Tejeda-Yeomans,
Phys.\ Lett.\ B {\bf 552} (2003) 48.

\bibitem{Catani:2003vu}
S.~Catani, D.~de Florian and G.~Rodrigo,
Phys.\ Lett.\ B {\bf 586} (2004) 323. 

\bibitem{Badger:2004uk}
Z.~Bern, L.~J.~Dixon and D.~A.~Kosower,
hep-ph/0404293;
S.~D.~Badger and E.~W.~N.~Glover,
hep-ph/0405236.



\bibitem{resex}
D.~de Florian and M.~Grazzini,
Phys.\ Rev.\ Lett.\  {\bf 85} (2000) 4678,
Nucl.\ Phys.\ B {\bf 616} (2001) 247;
S.~Catani, D.~de Florian and M.~Grazzini,
JHEP {\bf 0105} (2001) 025;
S.~Catani, D.~de Florian, M.~Grazzini and P.~Nason,
JHEP {\bf 0307} (2003) 028.

\bibitem{mcws}
CERN Workshop on {\it Monte Carlo tools for the LHC}, CERN, Geneva, July 2003.

\bibitem{Anastasiou:2003kj}
C.~Anastasiou, Z.~Bern, L.~J.~Dixon and D.~A.~Kosower,
Phys.\ Rev.\ Lett.\  {\bf 91} (2003) 251602.

\bibitem{Moch:2004pa}
S.~Moch, J.~A.~M.~Vermaseren and A.~Vogt,
Nucl.\ Phys.\ B {\bf 688} (2004) 101, 
hep-ph/0404111.

\bibitem{Kosower:2003np}
D.~A.~Kosower and P.~Uwer,
Nucl.\ Phys.\ B {\bf 674} (2003) 365.

\bibitem{Catani:2004nc}
S.~Catani, D.~de Florian, G.~Rodrigo and W.~Vogelsang,
hep-ph/0404240.

\bibitem{inprep}
S.~Catani, D.~de Florian, G.~Rodrigo and W.~Vogelsang,
in preparation.

\bibitem{Altarelli:1977zs}
G.~Altarelli and G.~Parisi,
Nucl.\ Phys.\ B {\bf 126} (1977) 298.


\bibitem{ellis} J.~R.~Ellis, Nucl.\ Phys.\ {\bf A684} (2001) 53.

\bibitem{Cao:1999da}
~G.~Cao and A.~I.~Signal,
Phys.\ Rev.\ D {\bf 60} (1999) 074021, and references therein.


\bibitem{Barone:1999yv}
V.~Barone, C.~Pascaud and F.~Zomer,
Eur.\ Phys.\ J.\ C {\bf 12} (2000) 243.

\bibitem{Portheault:2004xy}
B.~Portheault,
hep-ph/0406226.

\bibitem{Olness:2003wz}
F.~Olness {\it et al.},
hep-ph/0312323.


\bibitem{Zeller:2001hh}
G.~P.~Zeller {\it et al.}  [NuTeV Collaboration],
Phys.\ Rev.\ Lett.\  {\bf 88} (2002) 091802.

\bibitem{Davidson:2001ji}
S.~Davidson, S.~Forte, P.~Gambino, N.~Rius and A.~Strumia,
JHEP {\bf 0202} (2002) 037.

\bibitem{Kretzer:2003wy}
S.~Kretzer {\it et al.}, 
hep-ph/0312322.


\bibitem{fp}
W.~Furmanski and R.~Petronzio, Z.\ Phys.\ C {\bf 11} (1982) 293.

\bibitem{Catani:1994sq}
S.~Catani and F.~Hautmann,
Nucl.\ Phys.\ B {\bf 427} (1994) 475.

\bibitem{Gluck:1998xa}
M.~Gluck, E.~Reya and A.~Vogt,
Eur.\ Phys.\ J.\ C {\bf 5} (1998) 461.

\end{thebibliography}
\end{document}